\documentclass{iopart} 

\usepackage{latexsym}
\usepackage{iopams} 

\eqnobysec 
\newcommand{\odr}{\Or} 
\newcommand{\HGV}{\xi} 
\newcommand{\BGV}{\HGV_0} 
\newcommand{\CV}{\varsigma} 
\newcommand{\TAKV}{\eta_{\scriptscriptstyle (i)}} 
\newcommand{\TAKVJ}{\eta_{\scriptscriptstyle (j)}} 
\newcommand{\TAKVK}{\eta_{\scriptscriptstyle (k)}} 
\newcommand{\CIJK}{C_{\scriptscriptstyle (i) (j)}^{\scriptscriptstyle ~~~ (k)}} 
\newcommand{\potkil}{\lambda} 
\newcommand{\QED}{$\square$}
\newtheorem{definition}{Definition} 
\newtheorem{proposition}{Proposition} 
\newcommand{\Proof}{\textit{Proof}.~} 

\begin{document} 
\baselineskip .15in
\begin{flushright}
WU-AP/244/06
\end{flushright}

\title{\mbox{Universal properties from local geometric structure of Killing horizon}} 

\author{Jun-ichirou Koga} 

\address{Advanced Research Institute for Science and Engineering, 
Waseda University, Shinjuku, Tokyo 169-8555, Japan} 

\ead{koga@gravity.phys.waseda.ac.jp} 

\begin{abstract} 
We consider universal properties that arise from 
a local geometric structure of a Killing horizon. 
We first introduce a non-perturbative definition of such a local geometric structure,  which we call an asymptotic Killing horizon. 
It is shown that infinitely many asymptotic Killing horizons reside 
on a common null hypersurface, once there exists one asymptotic Killing horizon. 
The acceleration of the orbits of the vector that generates 
an asymptotic Killing horizon is then considered. 
We show that there exists 
the $\textit{diff}(S^1)$ or $\textit{diff}(R^1)$ sub-algebra 
on an asymptotic Killing horizon universally, which is picked out naturally 
based on the behavior of the acceleration. 
We also argue that the discrepancy between string theory and the Euclidean approach 
in the entropy of an extreme black hole may be resolved, 
if the microscopic states responsible for black hole thermodynamics 
are connected with asymptotic Killing horizons. 
\end{abstract}


\baselineskip 15pt
\section{Introduction} 
\label{sec:Introduction} 

The fact that a black hole behaves as a thermal object has been established  
\cite{BardeenCH73,Hawking75}
within the semi-classical theory of gravity, i.e., 
the quantum field theory on a \textit{classical} gravitational field, 
and then it is argued that the microscopic physics associated with 
black hole thermodynamics  
may be understood without resorting to details of quantum gravity 
(see, for example, \cite{Fursaev04,Carlip06a}). 
Indeed, within the semi-classical gravity, the entropy of 
a (non-extreme) black hole in Einstein gravity 
is shown to be given by one quarter of the area of the black hole horizon in Planck units, 
which is known as the Bekenstein--Hawking formula \cite{Hawking75,Bekenstein73}. 
Even in generalized theories of gravity, 
one can show that black hole entropy is expressed 
by geometric quantities on the black hole horizon \cite{WaldEntropy}. 
Furthermore, it has been shown \cite{JacobsonRevs,Jacobson95} 
that the thermal feature is possessed not only by a black hole horizon, 
but also by any causal horizons, 
including a cosmological horizon and an acceleration horizon, 
while the geometric structures far from the horizon differ between them.  
These facts may possibly suggest that the thermal feature of a black hole 
and the microscopic physics responsible for it are intimately 
connected with a universal and local geometric structure 
particular to presence of a horizon. 
However, it is quite difficult to understand, 
in terms of a universal and local geometric structure of a horizon, 
the statistical physics of the microscopic states of black hole thermodynamics.
Although the Euclidean approach to black hole thermodynamics 
\cite{GibbonsHawking77} gives the relation between 
the classical gravitational field of a black hole 
and the partition function of black hole thermodynamics, 
it does not clarify the nature of these microscopic states. 
The Euclidean approach shows only that there will exist the microscopic states 
associated with a single classical (on-shell) configuration in the phase space. 

One of the attempts to embark on this issue is 
to consider asymptotic symmetries on the horizon of an arbitrary black hole 
\cite{Carlips} 
(see \cite{Fursaev04,Carlip06a} for recent reviews and the references therein), 
following the success in the case of the B.T.Z. black hole 
whose entropy has been reproduced based on the asymptotic symmetries 
\textit{at infinity} \cite{Strominger98}. 
This attempt would be very interesting, if it worked 
in the same way as the case of the B.T.Z. black hole, 
since it would suggest that there is a sort of correspondence 
also between a black hole horizon and the conformal field theory, 
much like the AdS/CFT correspondence \cite{AdSCFT}. 
However, it seems that the fatal differences between 
the asymptotic symmetries on a horizon and those at infinity of 
the B.T.Z. black hole are obstructing the success of this attempt, 
and contrived methods with modifications in the framework 
have been proposed \cite{Carlip02,Carlip05} (see also \cite{KangKP04}). 
Without an established correspondence between a black hole horizon 
and the conformal field theory, however, it looks important to clarify, 
from a physical point of view,   
whether and how the asymptotic symmetries on a black hole horizon are 
related with black hole thermodynamics. 

Then, unless convincing physical arguments are provided,  
which prove that singularity or discontinuity of asymptotic symmetries 
is essential in black hole thermodynamics, 
it will be reasonable to focus on 
asymptotic symmetries that are regular and continuous on a black hole horizon. 
In a spherically symmetric spacetime with a Killing horizon, 
regular and continuous asymptotic symmetries 
described by asymptotic Killing vectors, 
along with a local structure of a Killing horizon, are analyzed previously \cite{Koga01}. 
However, assuming existence of a Killing horizon as a background is perturbative, 
and it actually spoils the idea of local structure 
of a horizon, because a Killing horizon is defined globally. 
Then, the first step in this paper is to define a local geometric structure of 
a Killing horizon non-perturbatively in a spacetime without any global isometries, 
by generalizing the notion of a Killing horizon and keeping its essential features. 
It will be interesting to analyze such a local geometric structure 
also from the viewpoint that to what extent the properties of 
a standard Killing horizon are robust and remain valid against such generalization. 
We will present in section \ref{sec:AsymKillH} 
the definition of such a local geometric structure, 
which we call an asymptotic Killing horizon, as well as the motivation for it 
and immediate consequences. 
In section \ref{sec:AsymSym}, we will consider the asymptotic Killing vectors 
on an asymptotic Killing horizon, and present the universal form of 
the asymptotic Killing vectors. 
We will also see that infinitely many asymptotic Killing horizons reside on 
a common null hypersurface, once there exists one asymptotic Killing horizon. 
In order to analyze the physical aspect of these asymptotic Killing horizons, 
we will introduce in section \ref{sec:SurfaceGravity} 
the notion of the surface gravity of an asymptotic Killing horizon 
and show the relation between the surface gravity and the timelike curves 
for which the asymptotic Killing horizon acts as a causal boundary, 
as if a Killing horizon does. 
The acceleration of these timelike curves will be analyzed 
in section \ref{sec:Acceleration}, based on which we will prospect   
the quantum features of asymptotic Killing horizons. 
The summary and discussions will be presented in section \ref{sec:discussion}. 

Throughout this paper, we will 
consider spacetimes with the Lorentzian signature and arbitrary dimensionality.  
All quantities in this paper are assumed to be smooth, 
while this assumption can be weakened appropriately.  

\section{Asymptotic Killing horizon} 
\label{sec:AsymKillH} 

We first consider how a local notion of a Killing horizon should be defined. 
A Killing horizon is a null hypersurface $H$ generated by the Killing vector $\chi^a$  
that becomes null on $H$ and tangent to $H$. 
By introducing a scalar function $u$ such that $u = 0$ on $H$, we thus have 
$\chi^a \chi_a = 0$ and $\chi^a \nabla_a u = 0$ on $H$, where $\chi^a$ of course 
satisfies the Killing equation 
${\cal L}_{\chi} g_{a b} = 0$ 
everywhere in the spacetime. 
However, we wish in this paper to focus only on the local geometry 
near the horizon, but not regions far away from it. 
What we require is then that the Lie derivative of the metric 
vanishes at least on the horizon, but may not away from the horizon. 

We emphasize also that it is not possible to define a local structure of a horizon 
only with a null hypersurface $H$ and the metric around $H$. 
It is necessary to specify an extra structure, 
such as the Killing vector $\chi^a$ in the case of a Killing horizon. 
Actually, one can perform a coordinate transformation from 
the flat metric to the one that looks  
like ``the asymptotic form of the Schwarzschild metric'' 
near the light cone emanating from an event in the Minkowski spacetime. 
However, the light cone in the Minkowski spacetime fails to be a horizon of any kind. 
In particular, it fails to be a Killing horizon, since there does not exist 
a Killing vector that is null on and tangent to the light cone, 
in sharp contrast with the Rindler horizon. 
Even in the cases where there do exist Killing vectors, we should specify 
whether and which Killing vector generates a Killing horizon. 
For example, the Rindler horizon behaves as a Killing horizon for 
a uniformly accelerated observer, while it does not for an inertial observer. 
These two observers are distinguished by the Killing vectors 
that are tangent to the orbits of the respective observers. 
Therefore, one needs to specify not only the null hypersurface, 
but also the vector that generates a local geometric structure of a Killing horizon. 

Based on these observations and 
by refining the previous perturbative definition \cite{Koga01}, 
we now define non-perturbatively an asymptotic Killing horizon, 
as a local generalization of the notion of a Killing horizon. 

\begin{definition} 
Let $H$ be a null hypersurface in a spacetime $( M , g_{a b} )$ 
equipped with the smooth metric $g_{a b}$. 
The pair $( H , \HGV^a )$ of $H$ and a smooth vector $\HGV^a$, 
which is not identically vanishing on $H$, is defined to be 
an asymptotic Killing horizon, and $\HGV^a$ is called the generator 
of the asymptotic Killing horizon $( H , \HGV^a )$, 
if there exists a neighborhood of $H$, in which   
\begin{enumerate} 
\item there exists a smooth scalar $u$, 
such that $u = 0$ and $\nabla_a u \neq 0$ on $H$. 
\item $\HGV^a$ satisfies the following three conditions: 
\begin{eqnarray} 
& & {\cal L}_{\HGV} g_{a b} = \odr(u) , 
\label{eqn:GenVecDef} \\ 
& & 
\HGV^a \HGV_a = \odr(u) ,  
\label{eqn:GenNull} \\ 
& & 
\HGV^a \nabla_{a} u = \odr(u) .  
\label{eqn:GenTngt} 
\end{eqnarray} 
\end{enumerate}
\label{def:AsymKillHor} 
\end{definition} 

The scalar $u$ in Definition \ref{def:AsymKillHor} 
is introduced to specify the null hypersurface $H$ 
in a regular and non-degenerate manner. 
In general, there will exist many scalar functions $u$ 
which satisfy $u = 0$ and $\nabla_a u \neq 0$ on $H$. 
However, it is easy to see that an asymptotic Killing horizon is defined 
without depending on the choice of $u$. 
We thus note that $u$ plays a similar role as 
the conformal factor in the conformal completion of an asymptotically 
flat or (anti-)de Sitter spacetime, whereas we need not conformally complete 
the spacetime in the present case. 
While (\ref{eqn:GenNull}) and (\ref{eqn:GenTngt}) take essentially the same forms 
as in the case of a Killing horizon, 
a key ingredient in Definition \ref{def:AsymKillHor} is (\ref{eqn:GenVecDef}), 
which states that the generator $\HGV^a$ behaves as if a Killing vector only on $H$, 
but  may not away from $H$. 

Since the normal $\nabla^a u$ to the null hypersurface $H$ 
of an asymptotic Killing horizon $( H , \HGV^a )$ does not vanish on $H$, 
and it is null on $H$ and tangent to $H$, 
$\nabla^a u$ must be proportional to the generator $\HGV^a$ on $H$.   
Then, there should exist a smooth scalar $n$ such that 
\begin{equation} 
 \HGV^a = n \nabla^a u + \odr(u) = \nabla^a ( n u ) + \odr(u) .   
\label{eqn:PropGen} 
\end{equation} 
By setting as $\potkil = n u$, we thus obtain 
\begin{equation} 
\HGV^a = \nabla^a \potkil + \odr(u) ,  
\label{eqn:PotKilDef} 
\end{equation} 
in the neighborhood of $H$. 
We will frequently use this smooth scalar $\potkil$ below, 
which we call, for simplicity, \textit{the potential of the generator} $\HGV^a$, 
due to the relation \eref{eqn:PotKilDef}, 
although it fails to play the role of a potential away from $H$. 
As we see from (\ref{eqn:PotKilDef}) and the fact that $\potkil$ vanishes 
on $H$, once $\HGV^a$ is given, 
$\potkil$ is uniquely determined up to $\odr(u^2)$  
without depending on the choice of $u$. 
Actually, if there exist two potentials $\potkil$ and $\potkil'$ of 
the same generator $\HGV^a$, we have 
$0 = \nabla_a ( \potkil - \potkil' ) + \odr(u)$, 
which shows $\potkil - \potkil' = \odr(u^2)$. 

When we consider any timelike vector $t^a$ that lies within the same 
half of the light cone as $\HGV^a$ does, 
we have $t^c \nabla_c \potkil = t_c \HGV^c < 0$ on $H$. 
Since $\potkil$ vanishes on $H$, 
we thus see that $\potkil$ necessarily changes its sign as we go across $H$ 
along $t^a$, and hence $H$ always partitions its neighborhood 
into the region of $\potkil > 0$ and that of $\potkil < 0$. 
If $\HGV^a$ is the future-pointing Killing vector that generates the Killing horizon 
of the Schwarzschild black hole, for example, 
the region of $\potkil > 0$ is the exterior region 
of the black hole, and the region of $\potkil < 0$ is the interior region. 
However, one should bear in mind that it is possible to change the sign of $\potkil$ 
by reversing the direction of $\HGV^a$. Therefore, $\potkil$ alone cannot  determine which side of the null hypersurface is exterior. 
One may recall that $\HGV^c \HGV_c$ also divides by its sign 
a neighborhood of the Killing horizon of a non-extreme black hole 
(a non-extreme Killing horizon, hereafter).  
However, it fails to do so in the case of 
an extreme black hole, 
since $\HGV^c \HGV_c$ is negative on the both sides of the Killing horizon 
of an extreme black hole (an extreme Killing horizon). 

As one may guess from the above features, 
$\potkil$ in the case of a standard Killing horizon  
is roughly proportional to $r - r_{\mathrm{H}}$ near the horizon,  
where $r$ is the standard radial coordinate 
and $r_{\mathrm{H}}$ is the horizon radius. 
On a standard Killing horizon, there exist not only the null Killing vector, 
but also spacelike Killing vectors, 
all of which are tangent to the horizon and commute 
with the null Killing vector, and the directional derivatives of $r - r_{\mathrm{H}}$ 
along these Killing vectors vanish. 
We can show that a similar but weaker property is possessed even by 
an asymptotic Killing horizon. 
Suppose that there exist, in a neighborhood of 
an asymptotic Killing horizon $( H , \HGV^a )$, 
the vectors $\CV^a$ which are tangent to $H$ as $\CV^a \nabla_a u = \odr(u)$ and 
``commute with $\HGV^a$'' in the sense that $\HGV_a$ (with the lowered index) 
is constant along $\CV^a$ on $H$ as ${\cal L}_{\CV} {\HGV}_a = \odr(u)$. 
By operating $\CV^a \nabla_a$ on the both sides of (\ref{eqn:PotKilDef}), 
we then obtain 
\begin{equation} 
\CV^a \nabla_a {\HGV}_b = \CV^a \nabla_a \nabla_b \potkil + \odr(u)  
= \nabla_b ( \CV^a \nabla_a \potkil ) - {\HGV}_a \nabla_b \CV^a + \odr(u) ,  
\end{equation} 
and hence 
\begin{equation} 
{\cal L}_{\CV} {\HGV}_b = 
\CV^a \nabla_a {\HGV}_b + {\HGV}_a \nabla_b \CV^a 
= \nabla_b ( \CV^a \nabla_a \potkil ) + \odr(u) .  
\label{eqn:XiConstSigma} 
\end{equation} 
Since we have ${\cal L}_{\CV} {\HGV}_b = \odr(u)$, 
(\ref{eqn:XiConstSigma}) gives $\nabla_b ( \CV^a \nabla_a \potkil ) = \odr(u)$,  
which implies   
\begin{equation} 
\CV^a \nabla_a \potkil = \odr(u^2) .  
\label{eqn:DerPotXi} 
\end{equation} 
In particular, the generator $\HGV^a$ of an asymptotic Killing horizon $( H , \HGV^a )$ 
is tangent to $H$, and satisfies ${\cal L}_{\HGV} {\HGV}_a = \odr(u)$. 
Therefore, we always have 
\begin{equation} 
\HGV^a \nabla_a \potkil = \odr(u^2) . 
\label{eqn:PotConstAll} 
\end{equation} 
In general, however, the vector $\CV^a$ need not be null, and can be spacelike.

\section{Asymptotic Killing vectors} 
\label{sec:AsymSym} 

The asymptotic Killing vectors were analyzed 
on a four-dimensional spherically symmetric Killing horizon \cite{Koga01} 
and on the Killing horizon of the four-dimensional Kerr--Newman black hole 
\cite{Koga-b}, 
which are shown to preserve the local structure of a Killing horizon. 
From a more general point of view, here we analyze asymptotic Killing vectors  
on an asymptotic Killing horizon, in order to extract their universal properties. 
Generalizing straightforwardly from the previous definition \cite{Koga01}, 
which is based on the idea that asymptotic Killing vectors generate 
the transformations of the associated asymptotic symmetry group, 
we thus define an asymptotic Killing vector on an asymptotic Killing horizon, 
as follows. 

\begin{definition} 
Let $( H , \HGV^a )$ be an asymptotic Killing horizon 
in a spacetime $( M , g_{a b} )$ equipped with the smooth metric $g_{a b}$. 
A smooth vector $\zeta^a$ is defined to be an asymptotic Killing vector 
on the asymptotic Killing horizon $( H , \HGV^a )$, 
if there exists a neighborhood of $H$, in which $\zeta^a$ satisfies 
\begin{equation} 
{\cal L}_{\zeta} \hat{g}_{a b} = \odr(u) , 
\label{eqn:AsymKilEq} 
\end{equation} 
for any smooth scalar $u$ such that $u = 0$ and $\nabla_a u \neq 0$ on $H$, 
where $\hat{g}_{a b}$ are arbitrary smooth metrics 
that coincide with $g_{a b}$ on $H$ 
as 
\begin{equation} 
\hat{g}_{a b} = g_{a b} + \odr(u) .  
\label{eqn:PerturbMetric} 
\end{equation} 
\label{def:AsymKillVec} 
\end{definition} 

One possible way to view the asymptotic Killing vectors $\zeta^a$ is that 
they generate the infinitesimal transformations of the asymptotic symmetry group, 
which map a metric $g_{a b}$ into $g_{a b} + {\cal L}_{\zeta} g_{a b}$ \cite{Koga01}. 
Since \eref{eqn:AsymKilEq} implies that 
the transformed metric deviates from the original one by $\odr(u)$, 
we introduce the equivalence class of metrics defined by \eref{eqn:PerturbMetric}. 
In order to ensure that these transformations close, 
we thus impose the condition that an asymptotic Killing vector 
$\zeta^a$ must satisfy (\ref{eqn:AsymKilEq}) for \textit{any} metrics $\hat{g}_{a b}$ 
in the equivalence class (\ref{eqn:PerturbMetric}).  
Otherwise, the successive transformations may 
take the metric out of the equivalence class, 
even if a vector $\zeta^a$ satisfies ${\cal L}_{\zeta} g_{a b} = \odr(u)$ for 
a specific metric $g_{a b}$. 
In contrast, it is sufficient for the generator $\HGV^a$ 
of an asymptotic Killing horizon $( H , \HGV^a )$ 
to solve (\ref{eqn:GenVecDef}) for the \textit{fixed} metric $g_{a b}$. 
In this sense, the condition on an asymptotic Killing vector is stronger than 
that on the generator of an asymptotic Killing horizon. 

It is not difficult to find an explicit form of asymptotic Killing vectors 
on an asymptotic Killing horizon, as we see from the following proposition. 
\begin{proposition} 
Let $( H , \HGV^a )$ be an asymptotic Killing horizon 
in a spacetime $( M , g_{a b} )$ equipped with the smooth metric $g_{a b}$, and     
let $\TAKV^a$ be smooth vectors, if any, which are spacelike on $H$ and satisfy 
\begin{equation} 
{\cal L}_{\TAKV} g_{a b} = \odr(\tilde{u}) , \qquad 
\TAKV^c \nabla_c \tilde{u} = \odr(\tilde{u}) , 
\label{eqn:XiITngt}  
\end{equation} 
in a neighborhood of $H$ 
for a smooth scalar $\tilde{u}$ 
such that $\tilde{u} = 0$ and $\nabla_a \tilde{u} \neq 0$ on $H$. 
Then, the vectors $\zeta^a$ defined by 
\begin{equation} 
\zeta^a = \omega \, \HGV^a 
+ a^{\scriptscriptstyle (i)} \, \TAKV^a 
- \potkil \, \nabla^a \omega + \odr(u^2) , 
\label{eqn:ZetaGenDef} 
\end{equation} 
for any smooth scalar $u$ such that $u = 0$ and $\nabla_a u \neq 0$ on $H$ 
are asymptotic Killing vectors on $( H , \HGV^a )$, 
where $\potkil$ is the potential of $\HGV^a$, 
$\omega$ is an arbitrary smooth scalar function, 
$a^{\scriptscriptstyle (i)}$ denotes arbitrary constants, 
and summation over the index $(i)$ is understood, if necessary. 
\label{prop:AsymKillVec}
\end{proposition} 
\Proof 
From \eref{eqn:XiITngt}, we immediately obtain  
\begin{equation} 
{\cal L}_{\TAKV} g_{a b} = \odr(u) ,  \qquad 
\TAKV^c \nabla_c u = \odr(u) , 
\label{eqn:XiIGenU} 
\end{equation} 
for any $u$, not only for $\tilde{u}$, as long as $u = 0$ and $\nabla_a u \neq 0$ on $H$.  
We then find from (\ref{eqn:GenTngt}) and (\ref{eqn:XiIGenU}) that 
the vectors $\zeta^a$ defined by (\ref{eqn:ZetaGenDef}) 
are tangent to $H$, i.e., 
\begin{equation} 
\zeta^c \nabla_c u = {\cal L}_{\zeta} u = \odr(u) , 
\label{eqn:ZetaTangent} 
\end{equation} 
for any $u$. 
By substituting (\ref{eqn:ZetaGenDef}), we write ${\cal L}_{\zeta} \hat{g}_{a b}$,    
for any $u$ and any $\hat{g}_{a b}$ of the form (\ref{eqn:PerturbMetric}), as 
\begin{equation} \fl 
{\cal L}_{\zeta} \hat{g}_{a b} 
= \omega {\cal L}_{\HGV} g_{a b} 
+ a^{\scriptscriptstyle (i)} {\cal L}_{\TAKV} g_{a b} 
+ {\HGV}_b \nabla_{a} \omega 
- ( \nabla_a \potkil ) ( \nabla_b \omega ) 
+ {\HGV}_a \nabla_b \omega 
- ( \nabla_b \potkil ) ( \nabla_a \omega ) + \odr(u) .  
\end{equation} 
From (\ref{eqn:GenVecDef}), (\ref{eqn:PotKilDef}), and (\ref{eqn:XiIGenU}), 
we obtain 
\begin{equation} 
{\cal L}_{\zeta} \hat{g}_{a b} = \odr(u) , 
\end{equation} 
and hence we see that 
$\zeta^a$ are asymptotic Killing vectors on $( H , \HGV^a )$. 
\QED 

In the explicit examples of a Killing horizon considered elsewhere 
\cite{Koga01,Koga-b}, 
the only possible form of asymptotic Killing vectors 
is shown to be given by (\ref{eqn:ZetaGenDef}), 
where one can see that the condition that 
asymptotic Killing vectors should satisfy (\ref{eqn:AsymKilEq}) 
for arbitrary $\hat{g}_{a b}$ 
restricts strongly the possible form of asymptotic Killing vectors. 
(The form of the asymptotic Killing vectors was presented 
in a primitive form in the previous paper \cite{Koga01}, 
but it is shown \cite{Koga-b} to be the same as (\ref{eqn:ZetaGenDef}).) 
A crucial feature of these asymptotic Killing vectors is that 
they contain the supertranslations in the null direction, 
i.e., the position-dependent translations 
described by the arbitrary function $\omega$. 
In a more general context, we now see from Proposition \ref{prop:AsymKillVec} 
that the asymptotic Killing vectors (\ref{eqn:ZetaGenDef}) are allowed on 
an arbitrary asymptotic Killing horizon. 
In particular, an arbitrary Killing horizon is an asymptotic Killing horizon, 
and hence the existence of the asymptotic Killing vectors (\ref{eqn:ZetaGenDef}) 
is universal on arbitrary Killing horizons. 

On the other hand, we have not proved that there do not exist 
other forms of asymptotic Killing vectors on a general asymptotic Killing horizon. 
However, since there are the cases \cite{Koga01,Koga-b} where 
the only possible form of asymptotic Killing vectors is given by 
(\ref{eqn:ZetaGenDef}), 
we see that the most general form of the asymptotic Killing vectors 
shared universally by arbitrary asymptotic Killing horizons is given  
by (\ref{eqn:ZetaGenDef}). 
By straightforward calculations using \eref{eqn:PotKilDef}, \eref{eqn:DerPotXi}, 
\eref{eqn:XiIGenU} and \eref{eqn:ZetaTangent}, 
one can also show that 
the asymptotic Killing vectors (\ref{eqn:ZetaGenDef}) form the Lie brackets algebra 
\begin{equation} 
\bigl[ \zeta_{1} , \zeta_{2} \bigr]^a 
= \omega_{12} \: \HGV^a
+ a_{12}^{\scriptscriptstyle (k)} \: \TAKVK^a
- \potkil \nabla^a \omega_{12} + \odr(u^2) , 
\label{eqn:LieBracketsFin} 
\end{equation} 
if $\TAKV^a$ commute with $\HGV^a$ on $H$ as 
$\left[ \HGV , \TAKV \right]^a = \odr(u^2)$ 
and form a sub-algebra on $H$ as 
$\left[ \TAKV , \TAKVJ \right]^a 
=  \CIJK \, \TAKVK^a + \odr(u^2)$, 
where the function $\omega_{12}$ and the constant $a_{12}^{\scriptscriptstyle (i)}$ 
are defined as 
\begin{equation} 
\omega_{12} \equiv \zeta^a_1 \, \nabla_a \omega_2  
- \zeta^a_2 \, \nabla_a \omega_1  ,  \qquad  
a_{12}^{\scriptscriptstyle (k)} 
\equiv a^{\scriptscriptstyle (i)}_1 \, a^{\scriptscriptstyle (j)}_2 \, \CIJK ,  
\label{eqn:StructureFuncConst} 
\end{equation} 
and the structure constant $\CIJK$ is truly constant, 
i.e., $\nabla_a \CIJK = 0$ everywhere. 

Now we consider a consequence that follows from 
Definition \ref{def:AsymKillHor} and Proposition \ref{prop:AsymKillVec}. 
Since the generator $\HGV^a$ of an arbitrary asymptotic Killing horizon $( H , \HGV^a )$  satisfies ${\cal L}_{\HGV} \hat{g}_{a b} = \odr(u)$ 
for any $\hat{g}_{a b}$ of the form (\ref{eqn:PerturbMetric}), 
due to (\ref{eqn:GenVecDef}) and (\ref{eqn:GenTngt}), 
we see that $\HGV^a$ is an asymptotic Killing vector on $( H , \HGV^a )$.  
Then, Proposition \ref{prop:AsymKillVec} states that 
the vectors $\tilde{\HGV}^a$ defined by 
\begin{equation} 
\tilde{\HGV}^a \equiv \omega \, \HGV^a - \potkil \, \nabla^a \omega + \odr(u^2) , 
\label{eqn:newGen} 
\end{equation} 
are asymptotic Killing vectors on $( H , \HGV^a )$, as well, 
which thus satisfy (\ref{eqn:GenVecDef}). 
We note also that $\tilde{\HGV}^a$ 
are null on $H$ and tangent to $H$, i.e., they satisfy 
(\ref{eqn:GenNull}) and (\ref{eqn:GenTngt}).  
Therefore, the vectors $\tilde{\HGV}^a$ are found to generate 
another asymptotic Killing horizons $( H , \tilde{\HGV}^a )$, 
as long as $\omega$ is not identically vanishing on $H$. 
Since $\omega$ is otherwise arbitrary, this fact indicates that 
once there exists an asymptotic Killing horizon $( H , \HGV^a )$, 
one can find infinitely many asymptotic Killing horizons $( H , \tilde{\HGV}^a )$ 
on the same null hypersurface $H$. 

It is then meaningful to consider   
the mapping $\pi_{\omega}$ from an asymptotic Killing horizon 
$( H , \HGV^a )$ to an another asymptotic Killing horizon $( H , \tilde{\HGV}^a )$,  
which is described as   
\begin{equation} 
\pi_{\omega} : ( H , \HGV^a ) \rightarrow ( H , \tilde{\HGV}^a ) , 
\label{eqn:GenMapDef} 
\end{equation} 
where $\tilde{\HGV}^a$ is given by (\ref{eqn:newGen}) 
and we now assume that $\omega$ is non-vanishing.
When we apply $\pi_{\omega}$ first and then $\pi_{\tilde{\omega}}$, 
an asymptotic Killing horizon $( H , \HGV^a )$ is mapped to $( H , \hat{\HGV}^a )$ 
through $( H , \tilde{\HGV}^a )$, where $\hat{\HGV}^a$ is defined by 
\begin{equation} 
\hat{\HGV}^a \equiv \tilde{\omega} \, \tilde{\HGV}^a 
- \tilde{\potkil} \, \nabla^a \tilde{\omega} + \odr(u^2) , 
\label{eqn:newnewGen} 
\end{equation} 
and $\tilde{\potkil}$ is the potential of $\tilde{\HGV}^a$. 
We find, by noting $\potkil = \odr(u)$, 
that $\tilde{\potkil}$ is related with $\potkil$ by 
\begin{equation} 
\tilde{\potkil} = \omega \, \potkil + \odr(u^2) , 
\label{eqn:PotKilTrans} 
\end{equation} 
since we have 
$\tilde{\HGV}^a = \nabla^a \tilde{\potkil} + \odr(u) 
= \omega \nabla^a \potkil + \odr(u)$, 
which should be consistent with (\ref{eqn:newGen}). 
From (\ref{eqn:newGen}), (\ref{eqn:newnewGen}), and (\ref{eqn:PotKilTrans}), 
we thus obtain  
\begin{equation} 
\hat{\HGV}^a 
= ( \tilde{\omega} \omega ) \, \HGV^a - \potkil \, \nabla^a ( \tilde{\omega} \omega ) 
+ \odr(u^2) , 
\end{equation} 
and find that the successive mappings defined by 
$\pi_{\tilde{\omega}} \circ \pi_{\omega}$ are described as 
\begin{equation} 
\pi_{\tilde{\omega}} \circ \pi_{\omega} = \pi_{( \tilde{\omega} \omega )} ,   
\label{eqn:MultMaps} 
\end{equation} 
which implies that 
the mappings (\ref{eqn:GenMapDef}) close and are associative. 
We also see from (\ref{eqn:MultMaps}) 
that the inverse of an arbitrary $\pi_{\omega}$ is given by $\pi_{\omega^{- 1}}$, 
and from \eref{eqn:newGen} that the identity is given by $\omega = 1$.  
Therefore, these transformations $\pi_{\omega}$ 
between asymptotic Killing horizons are shown 
to form a group. (The infinitesimal versions of these transformations 
are referred to as horizon deformations previously \cite{Koga01}.)   

We emphasize here that $\omega$ in (\ref{eqn:GenMapDef}) 
is an arbitrary non-vanishing function, and hence the restriction of 
the generator $\tilde{\HGV}^a$ to $H$ is an arbitrary null vector that is 
non-vanishing as long as $\HGV^a$ does not vanish.  
It is remarkable that multiplication of the generator $\HGV^a$ 
by an arbitrary function preserves the structure of an asymptotic Killing horizon, 
in contrast with the case of a Killing horizon, where 
only \textit{constant} rescaling of the Killing vector is allowed.  
Therefore, infinitely many asymptotic Killing horizons 
are degenerate on a Killing horizon universally, in the sense that 
all of these asymptotic Killing horizons 
possess the same local geometric structure and the transformations 
between them form a group, while
they are distinguished from each other because of the non-trivial rescaling by 
the arbitrary function $\omega$.   

\section{Surface gravity} 
\label{sec:SurfaceGravity} 

We now define the surface gravity $\kappa$ of 
an asymptotic Killing horizon $( H , \HGV^a )$ 
as the \textit{function} on $H$ that satisfies 
\begin{equation} 
\HGV^c \nabla_c \HGV^a = \kappa \, \HGV^a + \odr(u) , 
\label{eqn:SGAOrg} 
\end{equation} 
which is rewritten as 
\begin{equation} 
\nabla^a N^2 = 2 \, \kappa \, \HGV^a + \odr(u) , 
\label{eqn:NSqHGVRel} 
\end{equation} 
by defining $N^2 \equiv - \HGV^a \HGV_a$ 
and noting from (\ref{eqn:GenVecDef}) that we have 
$\HGV^c \nabla_c \HGV^a = \nabla^a N^2/ 2 + \odr(u)$. 
By using (\ref{eqn:PotKilDef}) and (\ref{eqn:NSqHGVRel}), 
we see that the surface gravity $\kappa$ 
of an asymptotic Killing horizon $( H , \HGV^a )$ 
gives the relation between $N^2$ and the potential $\potkil$ of $\HGV^a$ as 
\begin{equation} 
N^2 = 2 \kappa \potkil + \odr(u^2) .    
\label{eqn:NsqPotkilRel} 
\end{equation} 
Since $\HGV^a$ is orthogonal to $H$, we also have 
\begin{equation} 
\HGV_{[ a} \nabla_b \HGV_{c ]} = \odr(u) . 
\label{eqn:AsymHyperOrthg} 
\end{equation}  
By contracting (\ref{eqn:AsymHyperOrthg}) with $\nabla^a \HGV^b$ 
and using (\ref{eqn:GenVecDef}), we thus see that $\kappa$ is computed, 
in the same way as in the case of a Killing horizon, by 
\begin{equation} 
\kappa^2 = - \frac{1}{2} ( \nabla^a \HGV^b ) ( \nabla_a \HGV_b ) + \odr(u) .   
\label{eqn:SGASq} 
\end{equation} 

One should note, however, that the surface gravity $\kappa$ of 
an asymptotic Killing horizon $( H , \HGV^a )$ is not ensured to be constant 
in either the direction of $\HGV^a$ or the other directions on $H$. 
This is because the structure of an asymptotic Killing horizon 
is defined only locally, and it is not restrictive enough 
to compel the surface gravity into constant. 
Actually, when we operate $\HGV^c \nabla_c$ on the both sides of 
(\ref{eqn:NsqPotkilRel}) 
and use \eref{eqn:PotConstAll}, we obtain 
\begin{equation} 
\HGV^c \nabla_c N^2 = 2 \potkil \, \HGV^c \nabla_c \kappa + \odr(u^2) . 
\label{eqn:DerNSq} 
\end{equation} 
Since $\potkil$ is linear in $u$ to the leading order, 
the necessary and sufficient condition 
for $\HGV^c \nabla_c \kappa = 0$ is then found as 
\begin{equation} 
\HGV^c \nabla_c N^2 = \odr(u^2) . 
\label{eqn:KappaConstCond} 
\end{equation} 
If $\HGV^a$ satisfies the Killing equation globally, we have 
\begin{equation} 
\HGV^c \nabla_c N^2 = - \HGV^c \HGV^b {\cal L}_{\HGV} g_{c d} = 0 , 
\label{eqn:KilDerNsq} 
\end{equation} 
and thus (\ref{eqn:KappaConstCond}) is satisfied. However, this is not 
generally the case for the generator $\HGV^a$ of 
an asymptotic Killing horizon $( H , \HGV^a )$, since we have 
$\HGV^c \nabla_c N^2 = \odr(u)$. 
Therefore, for a generic asymptotic Killing horizon, 
(\ref{eqn:KappaConstCond}) is not satisfied, 
and hence the surface gravity is not constant in general. 

However, once there is an asymptotic Killing horizon $( H , \HGV^a )$, 
which does not necessarily satisfy (\ref{eqn:KappaConstCond}), 
we can always find an asymptotic Killing horizon that satisfies 
(\ref{eqn:KappaConstCond}), i.e., an asymptotic Killing horizon whose 
surface gravity is constant along its generator. 
To prove this, we consider an asymptotic Killing horizon $( H , \BGV^a )$,  
with the generator $\BGV^a$ written as 
\begin{equation} 
\BGV^a = \omega_0 \, \HGV^a - \potkil \, \nabla^a \omega_0 + \odr(u^2) , 
\label{eqn:RelToFindSpcl} 
\end{equation} 
and analyze whether there exists $\omega_0$ such that 
the surface gravity $\kappa_0$ of the asymptotic Killing horizon $( H , \BGV^a )$ 
satisfies $\BGV^c \nabla_c \kappa_0 = 0$, where $\kappa_0$ 
is shown, by using (\ref{eqn:SGAOrg}) and (\ref{eqn:RelToFindSpcl}), 
to be given as 
\begin{equation} 
\kappa_0 = \omega_0 \, \kappa + \HGV^c \nabla_c \omega_0 + \odr(u) .  
\label{eqn:TransEqToConstKappa} 
\end{equation} 
We note that (\ref{eqn:TransEqToConstKappa}) gives 
an ordinary differential equation on $H$ ($u = 0$). 
Actually, $\HGV^a$ is tangent to the null geodesics on $H$, 
while not necessarily affinely parametrized, and hence there exists 
a function $\varpi$ on $H$ such that $\HGV^a$ is written on $H$ as 
\begin{equation} 
\HGV^a = \varpi \left( \frac{\partial}{\partial v} \right)^a ,  
\label{eqn:VDefPre} 
\end{equation} 
where $v$ is the affine parameter of the null geodesics. 
By defining a new function $V$ by 
\begin{equation} 
V \equiv \int^v \varpi^{- 1} dv , 
\label{eqn:VDefTr} 
\end{equation} 
we then write (\ref{eqn:TransEqToConstKappa}) as 
\begin{equation} 
\frac{\partial \hat{\omega}_0}{\partial V} + \kappa \, \hat{\omega}_0 = \kappa_0 , 
\label{eqn:DEtoRefHor} 
\end{equation} 
where $\hat{\omega}_0$ is the restriction of $\omega_0$ to $H$, i.e.,   
$\omega_0 = \hat{\omega}_0 + \odr(u)$. Thus, the desired asymptotic Killing horizon 
$( H , \BGV^a )$ is obtained by solving (\ref{eqn:DEtoRefHor}), 
which is always possible.   

In the case where $\kappa$ vanishes identically on $H$, 
we find from (\ref{eqn:DEtoRefHor}) that $\hat{\omega}_0$ is given by 
\begin{equation} 
\hat{\omega}_0 = \kappa_0 V + C , 
\end{equation} 
where $C$ is an integration function on $H$ that does not depend on $V$.  
Hence, the vanishing surface gravity is transformed into either zero ($\kappa_0 = 0$) 
or non-zero ($\kappa_0 \neq 0$). 
In the case of $\kappa \neq 0$, on the other hand, 
the general solution of $\hat{\omega}_0$ is given as 
\begin{equation} 
\hat{\omega}_0 = \kappa_0 \: e^{- w} \int^{w} \frac{e^w}{\kappa} \: d w + C e^{- w} , 
\label{eqn:HatOm} 
\end{equation} 
where 
\begin{equation} 
w \equiv \int^V \kappa \: d V . 
\label{eqn:WDef} 
\end{equation} 
If $\kappa$ is non-vanishing and constant along $\HGV^a$, in particular, 
$w$ is found from (\ref{eqn:WDef}) as $w = \kappa V$, 
and hence $\hat{\omega}_0$ is given as 
\begin{equation} 
\hat{\omega}_0 = \frac{\kappa_0}{\kappa} + C e^{- \kappa V} . 
\label{eqn:OmegaKappaConst} 
\end{equation} 
(Here, an integration function, which may appear in $w$, is absorbed into $C$.) 
We see from (\ref{eqn:OmegaKappaConst}) that the particular solution, i.e., 
the solution with $C = 0$,  
has only the effect of multiplying the generator by a factor independent of $V$. 
Thus, the transformations between the asymptotic Killing horizons with 
the non-vanishing constant surface gravity essentially reduce to constant rescaling 
of the generators, which is possible even in the case of a Killing horizon. 
On the other hand, the homogenous solutions ($\kappa_0 = 0$) 
transform the surface gravity into zero, 
and then the resulting generator 
is tangent to the affinely parametrized null geodesics along $H$.  
We furthermore note that $\kappa_0$ can be chosen to be constant all over $H$, 
and hence that we can find an asymptotic Killing horizon whose surface gravity 
is an arbitrary constant all over $H$. 
In particular, on a non-extreme Killing horizon, 
there exists an asymptotic Killing horizon with the vanishing surface gravity, 
whose generator is tangent to the affinely parametrized null geodesics on the horizon. 
It is also possible to find asymptotic Killing horizons 
with the non-vanishing constant surface gravity on an extreme Killing horizon. 
We emphasize, however, that the surface gravity of an asymptotic Killing horizon 
is not necessarily related with the notion of temperature, in contrast with 
the surface gravity of a Killing horizon. 

Here we recall one of the important features of a Killing horizon. 
A Killing horizon is a causal boundary 
for the observers who stay at rest in the exterior region of the Killing horizon, 
i.e., fields propagating behind the Killing horizon cannot reach these observers. 
It is important also to note that the Killing vector 
which generates the Killing horizon 
is tangent to the orbits of these observers in the exterior region. 
Since an asymptotic Killing horizon is a local generalization of a Killing horizon, 
it is reasonable that these features 
are shared by asymptotic Killing horizons. 
We thus require that the orbits of the generator $\HGV^a$ of 
an asymptotic Killing horizon $( H , \HGV^a )$ should 
be timelike definitely in the exterior region, 
at least in the neighborhood where all the conditions for an asymptotic Killing horizon 
are satisfied. 
Then, one might consider that 
$N^2 = - \HGV^c \HGV_c$ is conditioned to be positive-definite 
in the region of $\potkil > 0$ of an asymptotic Killing horizon $( H , \HGV^a )$.  
However, it is actually sufficient to require here that $N^2$ is positive-definite 
\textit{on one side} of $H$. 
This is because even if $N^2$ is positive-definite in the region of $\potkil < 0$, 
we can find an asymptotic Killing horizon 
whose $N^2$ is positive-definite in the region of $\potkil > 0$, 
by simply reversing the direction of $\HGV^a$, 
which reverses the sign of $\potkil$ also. 
From now on, we thus assume that the directions of the generators $\HGV^a$ 
of all asymptotic Killing horizons $( H , \HGV^a )$ on a common null hypersurface $H$ 
have been chosen so that $\potkil > 0$ on a side of $H$ where $N^2 > 0$. 
Therefore, the functions $\omega$ in the transformations (\ref{eqn:GenMapDef}) 
are now restricted to be positive-definite, due to \eref{eqn:PotKilTrans}, 
and hence the generators of all asymptotic Killing horizons on $H$ point towards the same direction on $H$. We note also that the group structure of the mapping 
$\pi_{\omega}$ is preserved, even when $\omega$ are restricted to be 
positive-definite.  

With the condition $\potkil > 0$ on a side where $N^2 > 0$, 
we see from (\ref{eqn:NsqPotkilRel}) that 
the surface gravity of an asymptotic Killing horizon cannot be negative. 
If the surface gravity $\kappa$ is positive-definite, we indeed find that 
$N^2$ is ensured to be positive-definite in the region of $\potkil > 0$. 
However, it is subtle whether the surface gravity of an asymptotic Killing horizon 
can vanish. Actually, if we set $\kappa = 0$ in (\ref{eqn:NsqPotkilRel}), 
the sign of $N^2$ is not determined definitely. 
It occurs because the structure of an asymptotic Killing horizon 
determines the behavior of $\HGV^a$ no higher than $\odr(u)$.  
In the case of the extreme Killing horizon of the Kerr--Newman black hole 
($\kappa = 0$), however, 
$N^2$ is positive-definite in both of the exterior 
and the interior regions of the Killing horizon, which implies that 
$N^2$ is quadratic in $u$ to the leading order. 
Therefore, there does exist an asymptotic Killing horizon 
whose surface gravity vanishes. However, this is not ensured to be always the case, 
and the asymptotic Killing horizons with $\kappa = 0$ may fail, 
in general, to describe physically relevant circumstances we are interested in. 
Actually, one can show that 
the transformation (\ref{eqn:OmegaKappaConst}) with $\kappa_0 = 0$ 
and $C = const.$, 
which results in the vanishing surface gravity, 
transforms the Killing horizon of the Schwarzschild black hole 
into an asymptotic Killing horizon whose generator is null everywhere. 
Therefore, we will basically exclude asymptotic Killing horizons 
with the vanishing surface gravity from our considerations. 
However, an extreme Killing horizon \textit{is} physically relevant, 
and hence we will take into account extreme Killing horizons, 
where $N^2$ is assumed to be quadratic in $u$ to the leading order, 
so that $N^2 = 0$ has a double root at $u = 0$.

\section{Acceleration} 
\label{sec:Acceleration} 

The surface gravity of a Killing horizon scales in the same manner 
as the normalization of the Killing vector that generates the Killing horizon. 
This freedom of constant rescaling of the surface gravity is usually fixed, 
in the case of a Killing horizon,  
by determining the normalization of the Killing vector in a region far from 
the Killing horizon. When we focus only on a local geometric structure 
of a Killing horizon, we thus suffer from the ambiguity of this constant rescaling 
of the surface gravity.  
In the case of a Killing horizon with the non-vanishing surface gravity, however, 
the surface gravity of the Killing horizon is related with 
the acceleration of the orbits of the Killing vector. 
The acceleration is free from the ambiguity 
of the constant rescaling, and it is closely related with the quantum effect 
measured by a particle detector \cite{BirrellDavies82}. 

Then, we here consider the acceleration of the orbits of the generators $\HGV^a$ 
of asymptotic Killing horizons $( H , \HGV^a )$, 
whish are conditioned to be timelike, as we described in the previous section. 
The velocity $u^a$ of these orbits is given as  
$u^a \equiv \HGV^a / N$, and   
the magnitude $a$ of the acceleration $a^a = u^b \nabla_b u^a$ is computed as 
\begin{equation} 
a^2 \equiv a^c a_c 
= \frac{1}{N^4} \left[ ( \HGV^b \nabla_b \HGV^c ) ( \HGV^d \nabla_d \HGV_c ) 
+ \frac{1}{4 N^2} ( \HGV^b \nabla_b N^2 ) ( \HGV^d \nabla_d N^2 ) 
\right] . 
\label{eqn:AccelSqrGen} 
\end{equation} 
We note here that the second term in the bracket in (\ref{eqn:AccelSqrGen}) 
vanishes, if $\HGV^a$ satisfies the Killing equation globally, 
and \eref{eqn:AccelSqrGen} reduces in this case to 
\begin{equation} 
a^2  = \frac{1}{4} \frac{( \nabla_c N^2 ) ( \nabla^c N^2 )}{N^4} . 
\label{eqn:AccelMagSqrt} 
\end{equation} 
It is possible to rewrite (\ref{eqn:AccelSqrGen}) 
by following the same steps as in the case of a Killing horizon 
(see, e.g., \cite{Wald84}).  
By using (\ref{eqn:AsymHyperOrthg}) and (\ref{eqn:SGASq}), we actually have  
$( \HGV_b \nabla^b \HGV^c ) ( \HGV^a \nabla_a \HGV_c ) 
= N^2 \kappa^2 - \kappa \, \HGV^c \nabla_c N^2 + \odr(u^2)$,  
and then by substituting this relation into (\ref{eqn:AccelSqrGen}), 
we obtain  
\begin{equation} 
a^2 = \frac{1}{N^4} \left[ N^2 \kappa^2 - \kappa \, \HGV^c \nabla_c N^2 
+ \frac{1}{4 N^2} ( \HGV^b \nabla_b N^2 ) ( \HGV^c \nabla_c N^2 ) 
+ \odr(u^2) \right] . 
\label{eqn:AccelSqrMod} 
\end{equation} 

When the surface gravity $\kappa$ of an asymptotic Killing horizon $( H , \HGV^a )$ 
is vanishing ($\kappa = 0$) and constant along $\HGV^a$ ($\HGV^c \nabla_c \kappa = 0$), 
we have $N^2 = \odr(u^2)$ and $\HGV^c \nabla_c N^2 = \odr(u^2)$, 
and (\ref{eqn:AccelSqrMod}) reduces to 
\begin{equation} 
a^2  = \frac{1}{N^2} \left[ \frac{1}{4 N^4} ( \HGV^b \nabla_b N^2 ) ( \HGV^c \nabla_c N^2 ) 
+ \odr(u^0) \right] . 
\label{eqn:AccelExAsyKil} 
\end{equation} 
Thus, the acceleration $a$ may diverge as $u^{- 1}$ in this case. 
Actually, the value of $a^2$ is not determined definitely,  
and hence $a^2$ is not ensured to be positive-definite, 
since the second term in (\ref{eqn:AccelExAsyKil})  
is out of control of an asymptotic Killing horizon. 
In the case of an extreme Killing horizon, however, 
we see from \eref{eqn:KilDerNsq} and \eref{eqn:AccelMagSqrt} that 
$a^2$ is positive-definite, 
because $\HGV^a$ is timelike and $N^2$ is constant along $\HGV^a$. 
Furthermore, by writing as 
\begin{equation} 
N^2 = \alpha u^2 + \odr(u^3) , \qquad \HGV^a = n \nabla^a u + u Z^a , 
\end{equation} 
and noting $\HGV^a \HGV_a = - N^2 = \odr(u^2)$, 
we obtain  
\begin{equation} \fl 
\HGV^c \nabla_c N^2 
= \left[ \HGV^c \nabla_c \alpha - \frac{2 \alpha}{n} \HGV^c Z_c \right] u^2 
+ \odr(u^3) , \qquad 
( \nabla^c N^2 ) (\nabla_c N^2 ) 
= \frac{4 \alpha}{n} \left[ \HGV^c \nabla_c \alpha 
- \frac{2 \alpha}{n} \HGV^c Z_c \right] u^3 
+ \odr(u^4) . 
\end{equation} 
We thus find that the condition \eref{eqn:KilDerNsq} enforces 
\begin{equation} 
\HGV^c \nabla_c \alpha - \frac{2 \alpha}{n} \HGV^c Z_c = \odr(u) , 
\end{equation} 
which implies $( \nabla^c N^2 ) ( \nabla_c N^2 ) = \odr(u^4)$. 
Therefore, in the case of an extreme Killing horizon, we see, 
by using (\ref{eqn:AccelMagSqrt}) and recalling that $N^2$ is quadratic in $u$ 
to the leading order, 
that the acceleration $a$ is finite in the limit $u \rightarrow 0$, 
where the divergent terms in (\ref{eqn:AccelExAsyKil}) cancel out. 

On the other hand, in the case of $\kappa \neq 0$, 
where $N^2$ is linear in $u$ to the leading order, 
the acceleration $a$ is given, by using (\ref{eqn:NsqPotkilRel}), (\ref{eqn:DerNSq}), 
and (\ref{eqn:AccelSqrMod}), as     
\begin{equation} 
a = \frac{1}{N} 
\left[ \kappa - \frac{1}{2} \HGV^c \nabla_c \ln \kappa + \odr(u) \right] . 
\label{eqn:RelSGAclFin} 
\end{equation} 
When $\HGV^c \nabla_c \kappa = 0$, 
we see from (\ref{eqn:RelSGAclFin}) that  
the surface gravity $\kappa$ is related with the acceleration $a$ by 
\begin{equation} 
\kappa = \lim_{u \rightarrow 0} N a ,  
\label{eqn:SGAcclRelKil} 
\end{equation} 
as in the case of a Killing horizon. 
However, since $\HGV^c \nabla_c \kappa \neq 0$ generally, 
the acceleration $a$ receives the correction described by 
the second term in (\ref{eqn:RelSGAclFin}). 
We also see that, as in the case of a Killing horizon, 
the acceleration $a$ is free from the ambiguity of constant rescaling of $\HGV^a$, 
in contrast with the surface gravity $\kappa$. 
Although it diverges in the limit $u \rightarrow 0$, the physical origin 
of this divergent behavior is well understood, i.e., 
infinite amount of gravitational redshift, 
which is essential for a black hole to exhibit the thermal feature. 

In order to understand the behavior of the acceleration $a$, 
it is helpful to consider here the asymptotic Killing horizons $( H , \HGV^a )$ 
that reside on the same null hypersurface $H$ as 
a reference asymptotic Killing horizon $( H , \BGV^a )$ 
whose surface gravity $\kappa_0$ is constant along $\BGV^a$, 
and express the acceleration $a$ of the orbits of $\HGV^a$ 
in terms of the reference asymptotic Killing horizon $( H , \BGV^a )$. 
We then write $\HGV^a$ by using $\BGV^a$ as 
\begin{equation} 
\HGV^a = \omega \, \BGV^a - \potkil_0 \nabla^a \omega + \odr(u^2) , 
\label{eqn:AsyGentoRef} 
\end{equation} 
where $\potkil_0$ is the potential of $\BGV^a$, 
and we focus on the cases of $\kappa \neq 0$, as we mentioned above. 
Since $\kappa$ is shown, by the same calculation that leads to 
(\ref{eqn:TransEqToConstKappa}), to be related with $\kappa_0$ as 
\begin{equation}  
\kappa = \omega \left[ \kappa_0 + \BGV^c \nabla_c \ln \omega \right] 
+ \odr(u) ,   
\label{eqn:SGWRTGS} 
\end{equation} 
we have 
\begin{equation} 
\HGV^b \nabla_b \kappa 
= \omega^2 \left[ \kappa_0  \, ( \BGV^b \nabla_b \ln \omega ) 
+ ( \BGV^b \nabla_b \ln \omega ) ( \BGV^c \nabla_c \ln \omega ) 
+ ( \BGV^b \nabla_b \, \BGV^c \nabla_c \ln \omega ) \right] + \odr(u) . 
\label{eqn:KappaDerWRTRef} 
\end{equation} 
We notice that 
the acceleration $a$ in the limit of $u \rightarrow 0$ depends on $\omega$ 
only through its logarithmic derivative along $\BGV^a$. 
To see this clearly, we write as 
$\omega = \hat{\omega} + \odr(u)$, where $\hat{\omega}$ is 
the restriction of $\omega$ to $H$, 
and utilize the fact that 
the reference asymptotic Killing horizon $( H , \BGV^a )$ possesses 
the translational invariance along $\BGV^a$.  
Thus, although $\omega$ has been restricted to be positive-definite so far, 
we now analytically continue and set it to the Fourier modes as 
\begin{equation} 
\hat{\omega} = n_p \, f_p \: e^{i p V} . 
\label{eqn:ModeExpansion} 
\end{equation} 
Here,  $V$ is the scalar on $H$ defined by 
\begin{equation} 
\BGV^a = \left( \frac{\partial}{\partial V} \right)^a , 
\end{equation} 
in the same manner as (\ref{eqn:VDefPre}) and (\ref{eqn:VDefTr}), 
$n_p$ is an arbitrary normalization constant, 
$f_p$ is an arbitrary function on $H$ that does not depend on $V$, 
i.e., $\BGV^c \nabla_c f_p = 0$, and we call $p$ the wave number of 
the asymptotic Killing horizon. 
By noticing from (\ref{eqn:PotKilTrans}) that $\potkil$ 
is related with $\potkil_0$ as $\potkil = \omega \potkil_0$   
and using (\ref{eqn:NsqPotkilRel}), (\ref{eqn:RelSGAclFin}), 
(\ref{eqn:SGWRTGS}), and (\ref{eqn:KappaDerWRTRef}), 
we then obtain 
\begin{equation} 
a = \frac{1}{\sqrt{2 \potkil_0} \left( \kappa_0 + i p \right)^{1 / 2}} 
\left[ \left( \kappa_0 + i \frac{p}{2} \right) + \odr(u) \right] . 
\label{eqn:AccelMagFin} 
\end{equation} 

We see from (\ref{eqn:AccelMagFin}) 
that the acceleration $a$ in the limit $u \rightarrow 0$ 
depends only on the wave number $p$, 
and does not depend on how $\omega$ varies in the other directions. 
It is then worth considering the generators ${\HGV}^a_p$ in the representative class 
defined by 
\begin{equation} 
{\HGV}^a_p \equiv 
\omega_p \, \BGV^a - \potkil_0 \nabla^a \omega_p + \odr(u^2) , 
\end{equation} 
where $\omega_p$ is naturally ``normalized'' as 
\begin{equation} 
\omega_p \equiv e^{i p V} .  
\end{equation} 
From (\ref{eqn:LieBracketsFin}) and (\ref{eqn:StructureFuncConst}), 
we find that the Lie brackets between the generators ${\HGV}^a_p$ 
give rise to the sub-algebra 
\begin{equation} 
\left[ {\HGV}_p , {\HGV}_{p'} \right]^a = i \: ( p' - p ) \: {\HGV}^a_{p + p'} , 
\label{eqn:DiffCircle} 
\end{equation} 
which is isomorphic to the $\textit{diff} (S^1)$ or $\textit{diff}(R^1)$ algebra, 
depending on whether $p$ is discrete or continuous. 
Therefore, we see that there exists 
the $\textit{diff} (S^1)$ or $\textit{diff}(R^1)$ sub-algebra 
on an arbitrary asymptotic Killing horizon universally, 
and that this sub-algebra is picked out naturally and physically  
based on the behavior of the acceleration. 

We also find from (\ref{eqn:AccelMagFin}) that in the case of $\kappa_0 \neq 0$, 
the acceleration $a$ diverges as $u^{- 1 / 2}$ when the orbit approaches 
the null hypersurface $H$ of the asymptotic Killing horizons $( H , \HGV^a )$, 
as in the case of a non-extreme Killing horizon. 
In this case, (\ref{eqn:AccelMagFin}) reduces continuously to 
the standard relation (\ref{eqn:SGAcclRelKil}) for a Killing horizon 
in the limit of $p \rightarrow 0$. 
Whereas we are assuming $\kappa \neq 0$, we can consider 
the case of $\kappa_0 = 0$, as well, by considering that 
the reference asymptotic Killing horizon is an extreme Killing horizon.    
Also in this case, 
we see that $a$ diverges as $u^{- 1 / 2}$, while $p$ must be kept non-vanishing 
so that $\kappa \neq 0$. 
However, the acceleration of an extreme Killing horizon 
is finite, as we mentioned above. 
Therefore, the acceleration of the asymptotic Killing horizons 
on an extreme Killing horizon is discontinuous in the limit $p \rightarrow 0$. 
Although infinitely many asymptotic Killing horizons can reside on 
an extreme Killing horizon, these asymptotic Killing horizons, 
except for the extreme Killing horizon itself,  
behave in a manner completely different from the extreme Killing horizon. 
Rather, they behave similarly to non-extreme Killing horizons. 
In the limit of $p \rightarrow 0$, 
an extreme Killing horizon is thus clearly distinguished and isolated from 
both non-extreme Killing horizons and asymptotic Killing horizons on itself. 
In the limit of $p \rightarrow \infty$, on the other hand, 
the acceleration $a$ does not depend on $\kappa_0$. This indicates that, 
as $p$ increases, the behavior of 
the asymptotic Killing horizons on an extreme Killing horizon 
approaches that of the asymptotic Killing horizons on a non-extreme Killing horizon. 
In this sense, an extreme Killing horizon and a non-extreme Killing horizon will not be 
distinguished in the limit of $p \rightarrow \infty$. 

\section{Summary and Discussion} 
\label{sec:discussion} 

In order to analyze the local geometric structure of a Killing horizon, 
we first presented Definition \ref{def:AsymKillHor} of an asymptotic Killing horizon, 
by simply replacing the Killing equation with (\ref{eqn:GenVecDef}) 
and noting that such a local structure of a Killing horizon should be defined 
by the pair of the null hypersurface $H$ and the generator $\HGV^a$. 
We then defined asymptotic Killing vectors by 
Definition \ref{def:AsymKillVec}, 
and showed that the vectors (\ref{eqn:ZetaGenDef}) are always 
asymptotic Killing vectors on an arbitrary asymptotic Killing horizon. 
Since the only possible form of asymptotic Killing vectors 
on a standard Killing horizon is shown 
\cite{Koga01,Koga-b} to be given by (\ref{eqn:ZetaGenDef}), 
we saw that the most general form of the asymptotic Killing vectors 
that are universally possessed by arbitrary asymptotic Killing horizons 
is described by (\ref{eqn:ZetaGenDef}). 
As one can see from the derivations, 
the results in this paper strongly rely on the fact that the generator $\HGV^a$ 
of an asymptotic Killing horizon $( H , \HGV^a )$ is both tangent and normal to 
the hypersurface $H$, which is possible only on null hypersurfaces. 
Thus, the existence and the features of asymptotic Killing horizons 
shown in this paper will not be possible on timelike or spacelike hypersurfaces. 

A remarkable feature of asymptotic Killing horizons is 
that there exist infinitely many asymptotic Killing horizons $( H , \HGV^a )$ 
on a common null hypersurface.  
Each of these asymptotic Killing horizons is discriminated by its generator $\HGV^a$, 
which, in turn, is given by the asymptotic Killing vector that is null on $H$, 
but is actually an arbitrary null vector on $H$, since $\omega$ 
in (\ref{eqn:ZetaGenDef}) is an arbitrary function. 
In particular, infinitely many asymptotic Killing horizons reside 
on an arbitrary Killing horizon universally. 
We then showed that if the surface gravity of 
an asymptotic Killing horizon $( H , \HGV^a )$ is positive-definite, 
the orbits of the generator $\HGV^a$ are timelike 
in the exterior region, and hence that the null hypersurface $H$ behaves 
as a causal boundary for these timelike orbits, much like a Killing horizon. 
On the other hand, 
we excluded asymptotic Killing horizons with the vanishing surface gravity 
from our consideration, except for extreme Killing horizons, since they fail 
to act as a causal boundary. One might expect that such pathological behavior 
will be remedied by specifying higher order structures of an asymptotic Killing horizon. 
However, the analysis of the asymptotic Killing vectors on the cosmological horizon 
in the de Sitter spacetime \cite{Koga01} implies that 
non-trivial asymptotic Killing horizons will disappear and only the Killing horizon 
will survive, when higher order structures are specified. 

Since the acceleration of a timelike orbit is closely related with 
the quantum effect perceived by 
a particle detector running along the orbit, 
it may be possible to foresee the quantum feature of 
an asymptotic Killing horizon $( H , \HGV^a )$ 
from the behavior of the acceleration of the orbits of $\HGV^a$. 
By setting the function $\omega$ in \eref{eqn:AsyGentoRef} to 
the Fourier modes as \eref{eqn:ModeExpansion}, 
we then investigated the acceleration of these orbits. 
It was found that only the orbits of the generators $\HGV^a$ with  
different wave numbers $p$ possess different accelerations. 
In particular, the generators $\HGV^a_p$ whose $\omega$ are constant 
in the spatial directions on $H$ form 
the $\textit{diff}(S^1)$ or $\textit{diff}(R^1)$ sub-algebra (\ref{eqn:DiffCircle}). 
It is interesting that the asymptotic Killing horizons that are picked out 
based on such a physical argument 
are related with the algebra of the two-dimensional conformal field theory. 
One then may be interested in 
the classical central charge in the Poisson brackets algebra associated with 
these asymptotic Killing horizons, but this issue is 
discussed elsewhere \cite{Koga01,Koga-b}.  

We found from the divergent behavior of the acceleration 
that the difference between an extreme Killing horizon and a non-extreme Killing horizon stands out in the limit 
where the wave number $p$ of an asymptotic Killing horizon vanishes, 
and hence that an extreme Killing horizon looks isolated 
from non-extreme Killing horizons 
when the global structure of an extreme Killing horizon is respected. 
In the limit of $p \rightarrow \infty$, on the other hand, an extreme Killing horizon 
behaves in the same manner as a non-extreme Killing horizon. 
The latter will imply that an extreme black hole will not be 
distinguishable from a near-extreme black hole as far as its local structure 
of the horizon is concerned. 
This will be the case also 
in a high-energy theory that can probe the regime of $p \rightarrow \infty$, 
such as a quantum theory of gravity with  
a duality between the low-energy and the high-energy regimes. 
It is then interesting to recall here the discrepancy 
between string theory and the Euclidean approach
in the entropy of an extreme black hole.  
In string theory, which will be able to describe the regime of $p \rightarrow \infty$, 
the entropy of an extreme black hole is given by the same expression 
as that of a non-extreme black hole 
\cite{StromingerVafa96,Horowitz96,Sen05}. 
On the other hand, in the Euclidean approach, 
one evaluates the partition function only by a classical solution 
with respect for the global isometry of the Killing vector, 
which thus probes only the configuration corresponding to $p = 0$. 
Then, the entropy of an extreme black hole has been shown to vanish 
in the Euclidean approach, 
the expression for non-extreme black holes being inapplicable, 
and hence an extreme black hole is regarded as 
isolated from non-extreme black holes \cite{HawkingHR95,Teitelboim95}. 
Since asymptotic Killing horizons with the non-vanishing surface gravity 
exhibit the behavior consistent both of these 
reliable results in string theory and the Euclidean approach, 
this discrepancy in the entropy of an extreme black hole 
may be resolved, if the microscopic states of black hole thermodynamics 
are connected with the asymptotic Killing horizons. 
It is thus fascinating to explore further the quantum physics of 
asymptotic Killing horizons in future investigations. 

\ack 
I would like to thank R. M. Wald for suggesting a non-perturbative analysis 
of asymptotic Killing horizons, which has motivated me to pursue this work. 
I am grateful also to A. Hosoya, K. Maeda, M. Natsuume, 
G. Kang and M. Park for useful discussions and encouragement. 


\end{document}